\documentclass{jpsj2}
\newcommand{\pos}{PrOs$_4$Sb$_{12}$}

\def\runtitle{Magnetic Phase Diagram of the Heavy-Fermion Superconductor \pos}
\def\runauthor{Tayama \textit{et al.}}

\title{%
Magnetic phase diagram of the heavy fermion superconductor \pos
}

\author{%
Takashi TAYAMA$^1$\thanks{E-mail address: tayama@issp.u-tokyo.ac.jp}, Toshiro SAKAKIBARA$^1$, \\Hitoshi SUGAWARA$^2$, Yuji AOKI$^2$, and Hideyuki SATO$^2$
}

\inst{%
$^1$Institute for Solid State Physics, University of Tokyo, Kashiwa, Chiba 277-8581\\
$^2$Department of Physics, Tokyo Metropolitan University, Minami-Ohsawa 1-1, Hachioji, Tokyo 192-0397
}

\recdate{\today}

\abst{%
We investigated the magnetic phase diagram of the first Pr-based heavy fermion superconductor \pos\ by means of high-resolution dc magnetization measurements in low temperatures down to 0.06~K. The temperature dependence of the magnetization $M(T)$ at 0.1~kOe exhibits two distinct anomalies at $T_\mathrm{c1}=1.83$~K and $T_\mathrm{c2}=1.65$~K, in
agreement with the specific heat measurements at zero field. Increasing magnetic field $H$, both $T_\mathrm{c1}(H)$ and $T_\mathrm{c2}(H)$ move toward lower temperatures without showing a tendency of intersecting to each other. Above 10~kOe, the transition at $T_\mathrm{c2}(H)$ appears to merge into a line of the peak effect which is observed near the upper critical field $H_\mathrm{c2}$ in the isothermal $M(H)$ curves, suggesting a common origin for these two phenomena. The presence of the field-induced ordered phase (called phase A here) is confirmed for three principal directions above 40~kOe, with the anisotropic A-phase transition temperature $T_\mathrm{A}$: $T_\mathrm{A}^{[100]} > T_\mathrm{A}^{[111]} >T_\mathrm{A}^{[110]}$.  The present results are discussed on the basis of crystalline-electrical-field level schemes with a non-magnetic ground state, with emphasis on a $\Gamma_1$ singlet as the possible ground state of Pr$^{3+}$ in \pos.
}

\kword{\pos, heavy-fermion superconductor, mixed state, magnetization, phase diagram, crystalline electric field effect
}

\begin{document}
\sloppy
\maketitle

\section{Introduction}
The filled skutterudite compound \pos\ is considered to be the first Pr-based heavy fermion (hf) superconductor.
The presence of the heavy quasiparticles has been suggested by the large specific heat jump $\Delta C/\gamma T_\mathrm{c}=500$~mJ/K$^2$mol at $T_\mathrm{c}=1.85$~K \cite{Bauer02}, and more directly by the recent de Haas-van Alphen experiments \cite{Sugawara02PRB}.
\pos\ has received considerable attention, not only because it is the first Pr-based hf superconductor but also due to its unusual superconducting (sc) properties. Absence of a coherence peak in the temperature variation of the nuclear spin-lattice relaxation rate in the Sb-NQR experiment is suggestive of non s-wave superconductivity \cite{Kotegawa02}. However, it is under debate whether the sc gap has nodes or not. The thermal conductivity measurement in rotating magnetic fields has reported that the sc energy gap has point nodes \cite{Izawa02}. On the contrary, the muon-spin relaxation \cite{MacLaughlin02} and the NQR \cite{Kotegawa02} experiments on \pos\ have reported an isotropic superconducting (sc) energy gap. 

Unconventional superconductivity in \pos\ is also suggested by its unusual phase diagram. The recent specific heat measurements at zero field revealed double sc transitions at $T_\mathrm{c1}=1.75$~K and $T_\mathrm{c2}=1.85$~K \cite{Maple02,Vollmer02}. Moreover, the thermal conductivity measurement has claimed that the gap symmetry changes from two-fold to four-fold as the field increases \cite{Izawa02}. Both experiments suggest the presence of the multiple sc phase in \pos\ \cite{Izawa02,Maple02,Vollmer02}. However, this interesting possibility certainly needs to be verified by various experimental methods.

One of the important issues in \pos\ is how the hf state is formed. In this regard, a two-channel quadrupolar Kondo effect has been proposed on the assumption that  the CEF ground state for trivalent Pr ion is a non-Kramers $\Gamma_3$ doublet \cite{Cox87}. On the contrary, Aoki \textit{et al.} reported that the specific heat data of \pos\ are better described by a $\Gamma_1$ ground state model \cite{Aoki02}.  Thus, the CEF ground state in \pos\ has not been established yet. In this respect, it would be important to notice that the recent specific heat measurements on \pos\ revealed a field-induced ordered phase in high fields above 40~kOe for the [100] direction \cite{Aoki02}. Since such a field-induced ordering has never been observed in the Ce-based hf systems, clarification of its origin would be a clue to resolve the electronic state in this compound.

From the above point of view, we investigated the sc state and the magnetic phase diagram of \pos\ in detail by dc magnetization measurements. The temperature dependence of the magnetization $M(T)$ is found to exhibit the double sc transition in magnetic fields up to 14 kOe. In particular, the presence of the A-phase is found not only for [100] direction but also for all the principal directions. 

The paper is organized as follows. After a brief description of the experiment in Sec.~II, we show the magnetization results in Sec.~III. In Sec.~IV an analysis based on the CEF model is given, followed by a discussion on the sc phase and phase A in Sec.~V. The paper closes with a summary in Sec.~VI.

\section{Experimental}
A single crystal of PrOs$_4$Sb$_{12}$ (20~mg) was grown by the Sb-self-flux method \cite{Takeda99,Bauer01}, using high-purity raw materials 4N (99.99\% pure)-Pr, 4N-Os and 6N-Sb. 
For the dc magnetization measurement below 2 K, we used a capacitive Faraday force magnetometer installed in a $^3$He-$^4$He dilution refrigerator \cite{Sakakibara94}. The resolution of the system is better than 10$^{-5}$~emu. Throughout the measurements, we applied a field gradient (1~kOe/cm) which is indispensable to this method. The resulting field distribution inside the sample was estimated to be less than 150 Oe. The magnetic field was applied along the three principal directions of the body centered cubic PrOs$_4$Sb$_{12}$ structure: [100], [110], and [111]. A SQUID magnetometer (MPMS, Quantum Design Co.) was also used to measure the dc magnetization in the temperature range of 2-300~K in field up to 70~kOe.

\section{Results}

\begin{figure} 
\begin{center}
\includegraphics[scale=0.45]{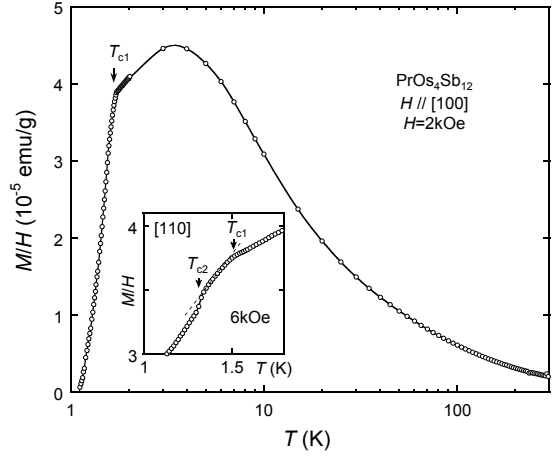}
\end{center}
\caption{Magnetic susceptibility $M(T)/H (\equiv \chi (T))$ of PrOs$_4$Sb$_{12}$ in a magnetic field of 2~kOe applied along the [100] direction. The arrow indicates the superconducting transition temperature $T_\mathrm{c1}$. The inset demonstrates the double transition at $T_\mathrm{c1}=1.51$~K and $T_\mathrm{c2}=1.31$~K in the $M(T)/H$ curve for $H \parallel [110]$ at $H=6$~kOe.}
\end{figure}
Figure~1 shows the temperature dependence of the magnetic susceptibility $M(T)/H (\equiv \chi (T))$ of PrOs$_4$Sb$_{12}$ in a fixed field of 2~kOe applied along the [100] direction, plotted on a semi-logarithmic scale. Above 50~K, the $\chi (T)$ curve follows a Curie-Weiss law with the effective moment $\mu_\mathrm{eff}$=3.5~$\mu_\mathrm{B}$ and the Weiss temperature $\theta_\mathrm{CW}$=-15~K. The obtained values of $\mu_\mathrm{eff}$ and $\theta_\mathrm{CW}$ are in good agreement with the previously reported ones \cite{Bauer02}.  The $\chi (T)$ curve shows a rather broad peak around 3.5~K as reported before \cite{Bauer02}, though the origin is not clarified yet. On further cooling, $\chi$ continues decreasing until a sharp kink due to the superconductivity appears at 1.51~K.

\begin{figure} 
\begin{center}
\includegraphics[scale=0.45]{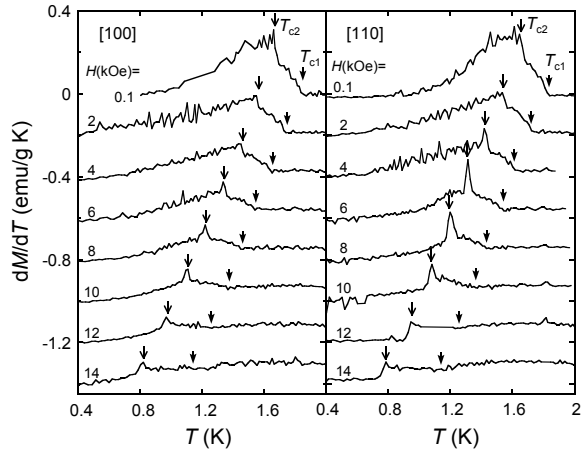}
\end{center}
\caption{Temperature derivative of the magnetization $\mathrm{d} M/\mathrm{d} T$ plotted as a function of $T$ at several fields below 14~kOe for two field orientations. The data were taken upon gradually warming the sample after field cooling. Two types of arrow indicate the double transition at $T_\mathrm{c1}$ and $T_\mathrm{c2}$.}
\end{figure}
Double sc transition at $T_\mathrm{c1}=1.85$~K and $T_\mathrm{c2}$=1.75~K in zero field has been reported in the specific heat $C(T)$ measurements of \pos\ \cite{Maple02,Vollmer02}.
To check the presence of the double transition, we carefully measured the temperature dependence of the magnetization in low fields below 14~kOe. Temperature derivatives of the magnetization $\mathrm{d} M(T)/\mathrm{d} T$ at several fields for $H \parallel [100]$ and [110] are shown in Fig. 2. The data were taken upon gradually warming the sample after cooling in magnetic field from above $T_\mathrm{c1}$.
In the $\mathrm{d} M(T)/\mathrm{d} T$ curve at $H=100$ Oe, we found double transition at $T_\mathrm{c1}=1.83$~K and $T_\mathrm{c2}=1.65$~K, which are close to those determined in the $C(T)$ measurements at zero field.  In the $C(T)$ measurements the double transition was not visible at  5~kOe \cite{Maple02,Vollmer02}. Here we were able to trace the two anomalies in d$M$/d$T$ up to 14~kOe, as seen in Fig.~2. With increasing the field, both anomalies shift toward the lower temperature side. Interestingly, the anomaly at $T_\mathrm{c1}$ becomes weaker, whereas the one at $T_\mathrm{c2}$ becomes more pronounced.  The inset of Fig.~1 shows the double transition at 6~kOe in the $M(T)/H$ curve for $H \parallel [110]$, in which a small but distinct magnetization jump can be seen at $T_\mathrm{c2}$.

\begin{figure} 
\begin{center}
\includegraphics[scale=0.45]{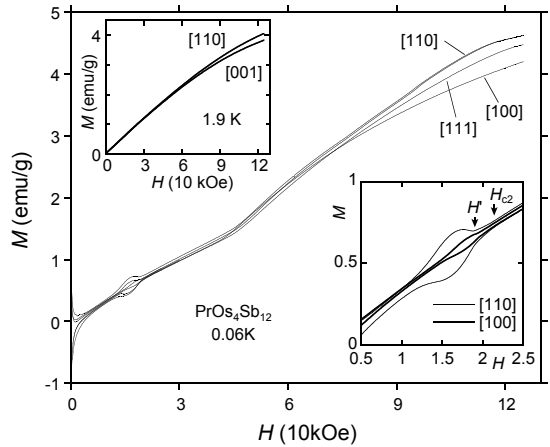}
\end{center}
\caption{Isothermal magnetization curves of PrOs$_4$Sb$_{12}$ at the base temperature of 0.06~K for three principal directions [100], [110], and [111]. The inset shows the $M(H)$ curve for [100] and [110] direction around the peak effect.  Two distinct anomalies at $H'$ and $H_\mathrm{c2}$ are shown in the inset.}
\end{figure}
The magnetization process $M(H)$ at 1.9~K, which is well above the ordering temperatures, are shown in the upper inset of Fig.~3.  The magnetization is nearly isotropic. The difference between the $M$ values for two directions is only 5~\% even at the high field of 125~kOe, with the easy magnetic direction being [110]. 
The $M(H)$ curves of \pos\ at the base temperature of 60~mK for three principal directions [100], [110], and [111]  are shown in Fig.~3. The magnetic anisotropy remains small even in the base temperature of 0.06~K.
In the $M(H)$ curves, superconductivity manifests itself in the hysteresis at low fields.
A marked peak effect is seen around 15~kOe for all the directions. As seen in the lower inset of Fig.~3, the peak effect is considerably anisotropic; the peak effect in the [110] direction becomes most pronounced. 

In Fig.~4 we plot the differential susceptibility  $\mathrm{d} M/\mathrm{d} H$ for the increasing-field process for three principal directions. The $\mathrm{d}M/\mathrm{d}H$ curve for [100] direction at 0.06~K in the vicinity of $H_\mathrm{c2}$ is shown in the inset of Fig.~4. Here we define the upper critical field $H_\mathrm{c2}$ by a sharp kink in the $\mathrm{d}M/\mathrm{d}H$ curve as indicated in Fig.~4. The so-derived $H_\mathrm{c2}^0\equiv H_\mathrm{c2}(T=0.06$~K) for $H \parallel [100]$ is  22.4~kOe, which coincides well with $H_\mathrm{c2}^0$ determined from the resistivity measurements \cite{Bauer02}. $H_\mathrm{c2}^0$ for $H \parallel [100]$ is 1.4~\% higher than that (=22.1~kOe) for $H \parallel [110]$ and [111].  
When the field is increased, the sharp structures due to the peak effect appear below $H_\mathrm{c2}$. The
onset of the peak effect is characterized by a dip in d$M$/d$H$, as usual.
Interestingly, the peak effect abruptly diminishes at the field denoted by $H'$
where a discontinuous jump is observed in d$M$/d$H$. This is a new feature that
has not been observed in the conventional peak effect, suggesting a certain
dramatic change in the vortex state.

\begin{figure} 
\begin{center}
\includegraphics[scale=0.45]{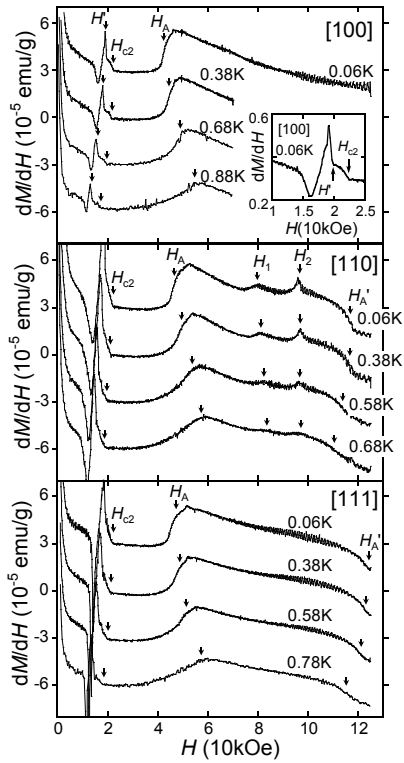}
\end{center}
\caption{Differential susceptibility $\mathrm{d} M(H)/\mathrm{d} H$ for the increasing field process at several temperatures below 0.88~K. The results for the three principal directions [100], [110], and [111] are shown. For clarity, the origin of each curve is shifted vertically. The arrows indicate various transition fields. The inset shows the $\mathrm{d} M/\mathrm{d} H$ in the vicinity of  $H_\mathrm{c2}$ for $H \parallel [100]$.}
\end{figure}
Going back to the data in Fig.~3, a clear upward bend of $M(H)$ is visible around 45~kOe in the normal state for all the directions, indicating that a second-order phase transition sets in. The present result is consistent with the previous $C(T)$ measurements which revealed the presence of the field-induced ordered phase (called phase A here) for $H \parallel [100]$ \cite{Aoki02}. 
The $\mathrm{d} M/\mathrm{d} H$ curves in Fig.~4 show the A-phase transition more clearly.
For $H \parallel [100]$ only a single anomaly is visible at $H_\mathrm{A}$=43~kOe above $H_\mathrm{c2}$, which is defined as the transition field from the normal state to the A-phase. This anomaly gradually shifts towards the high-field side as the temperature increases. For $H \parallel [110]$ four distinct anomalies are found at 47, 81, 97, and 116~kOe in the $\mathrm{d} M/\mathrm{d} H$ curve at 0.06~K. From the lower field side we assign the four critical fields as $H_\mathrm{A}$, $H_1$, $H_2$, and $H_\mathrm{A}'$, respectively.
$H_\mathrm{A}'$ is considered to be the field where the A-phase disappears, whereas $H_1$ and $H_2$ probably come from certain change in the ordered structure.
As the anomaly at $H_2$ is markedly sharp, this phase transition is likely of first order.
With increasing the temperature, $H_1$ and $H_2$ do not shift very much, whereas $H_\mathrm{A}$ and $H_\mathrm{A}'$ approach to each other gradually.
For $H \parallel [111]$ there are two anomalies at $H_\mathrm{A}$=45~kOe and $H_\mathrm{A}'$=124~kOe in the $\mathrm{d} M/\mathrm{d} H$ curve at 0.06~K. These anomalies likewise approach to each other as the temperature increases. The de Haas-van Alphen (dHvA) effect is clearly seen above 80~kOe for all the directions, which is indicative of the excellent quality of our sample. The dHvA frequency is estimated to be about $1.03 \times 10^3$~T, which coincides well with that of the $\beta$ branch obtained by the field modulation technique \cite{Sugawara02PRB}.

\begin{figure} 
\begin{center}
\includegraphics[scale=0.45]{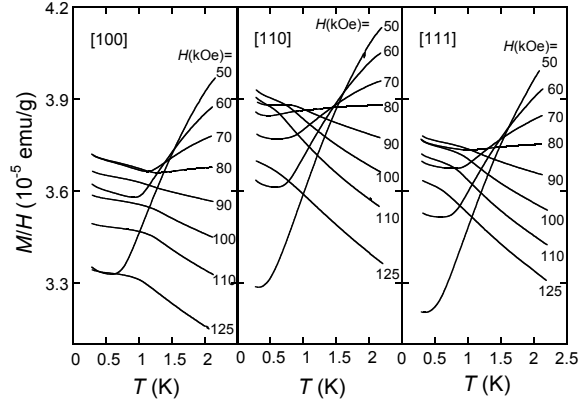}
\end{center}
\caption{Temperature dependence of the $M/H$ value at several fields above 50~kOe for three principal directions [100], [110], and [111].}
\end{figure}
\begin{figure} 
\begin{center}
\includegraphics[scale=0.45]{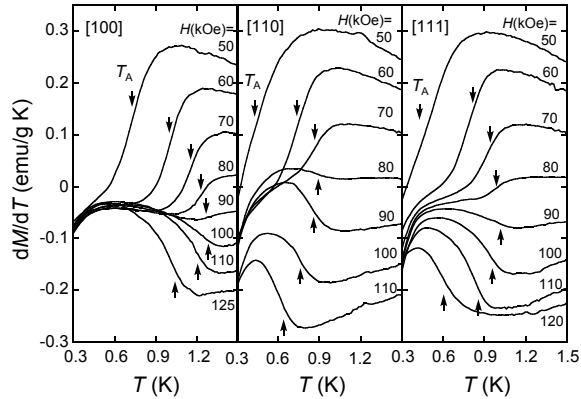}
\end{center}
\caption{Temperature derivative of the magnetization $\mathrm{d} M/\mathrm{d} T$ at several fields above 50~kOe for three different directions [100], [110], and [111]. The arrows indicate the A-phase transition temperature $T_\mathrm{A}$.}
\end{figure}
We next move on to the temperature dependence of the magnetization $M/H$. Figure~5 shows the $M(H)/H$ curves at several fields above 50~kOe for three principal directions. As already shown in Fig.~1, $M/H$ at low field exhibits a broad maximum around 3.5~K. This feature can still be seen in the $M/H$ data for 50~kOe which show a strong increase above 1~K. 
One can see the A-phase transition as a sudden change of the slope in the $M(T)/H$ curve at around 0.7~K.
To show the A-phase transition more clearly, the temperature derivatives of the magnetization  $\mathrm{d}M/\mathrm{d}T$ are shown in Fig.~6. We can easily determine the A-phase transition temperature $T_\mathrm{A}$ by the position of a step in the d$M$/d$T$ data as shown by the arrows. $T_\mathrm{A}$ first increases with $H$, taking the maximum at around 90~kOe, and then decreases at
higher fields.
Although not clearly seen in Fig.~5 due to a limited temperature
window of the present measurements, the broad maximum in $M/H$ appearing at
low field  (Fig.~1) gradually shifts to the low field side as $H$ increases.
The maximum eventually disappears at around $H \sim 90$ kOe, above which $M/H$ becomes a
decreasing function of $T$. 
It should be noted that the field where the broad peak in the $M(T)$ curve disappears is close to the field where $T_\mathrm{A}$ is the highest. The observation suggests that the origin of the maximum in $\chi$ is closely related to the existence of phase A.

\begin{figure}
\begin{center}
\includegraphics[scale=0.45]{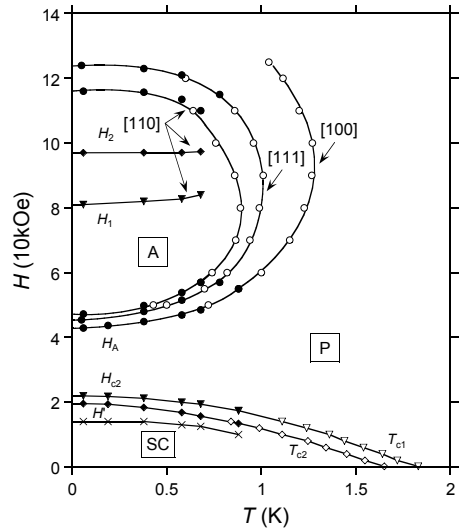}
\end{center}
\caption{$H-T$ phase diagram of PrOs$_4$Sb$_{12}$. A, P, and SC represent for the A phase, paramagnetic phase, and the superconducting phase, respectively. Open and closed symbols were determined by the $\mathrm{d} M(T)/\mathrm{d} T$ and  $\mathrm{d} M(H)/\mathrm{d} H$ data, respectively. Since the superconducting phase is nearly isotropic on this scale, only the data for $H \parallel [100]$ is shown in the figure. Cross symbols denote to the onset of the peak effect in the $M(H)$ process.}
\end{figure}
The present magnetization data are summarized in the $H-T$ phase diagram displayed in Fig.~7.
Open and closed symbols are the transition points derived from the $M(T)$ and $M(H)$ data, respectively. $T_\mathrm{c1}$ and $T_\mathrm{c2}$ are nearly isotropic. Since $T_\mathrm{c2}(H)$ smoothly connects to $H'$, the origin of $T_\mathrm{c2}$ is presumably the same as the  peak effect observed in the $M(H)$ curve. Recently, a multiple sc phase diagram has also been reported from the thermal conductivity measurements on \pos\ \cite{Izawa02}. 
They observed that a gap symmetry changes from two-fold to four-fold at a
certain field. Its boundary, beginning near $T_\mathrm{c2}(0)$, increases up to ~0.7~T at about 0.5~K with decreasing $T$. The boundary $T_\mathrm{c2}(H)$ determined in the present experiment is, however, essentially different from that reported in ref.~\cite{Izawa02}. So far, we have not succeeded in finding any appreciable anomaly in $M(H)$ around 0.7~T as reported in ref.~\cite{Izawa02}.

Regarding the field-induced ordering, it should be emphasized that the A-phase exists for all the directions. Similar results have also been reported independently in ref.~\cite{Tenya02}. We found that the anisotropy of the ordering temperature $T_\mathrm{A}$ is $T_\mathrm{A}^{[100]}(H) > T_\mathrm{A}^{[111]}(H) >T_\mathrm{A}^{[110]}(H)$. Due to the unique phase diagram, it has been proposed that the A-phase is an antiferro-quadrupolar (AFQ) state \cite{Aoki02,Maple02}. While this is feasible, however, we cannot identify the order parameter only from the shape of the phase diagram, as will be shown later. We note that the anisotropy of $T_\mathrm{A}$ in \pos\ is different from that of the AFQ transition temperature $T_\mathrm{Q}$ in the $O_\mathrm{xy}$-type AFQ compound CeB$_6$ \cite{Hiroi97} and the $O_2^0$-type AFQ compound PrPb$_3$ \cite{Tayama01}.

\section{Analysis}
In this section we analyze the thermodynamic properties of \pos\ by a CEF model. We believe that the localized $f$-electron picture ($4f^2$) is a good starting point to understand the magnetic properties of the system \cite{Sugawara02PRB}, though the hybridization effect between the $f$ electrons and the conduction electrons is surely important. We introduce the following mean field Hamiltonian:

\begin{eqnarray}
H=H_\mathrm{CEF}- g_\mathrm{J}\mu_\mathrm{B}\boldsymbol{J}\cdot (\boldsymbol{H}+n \langle \boldsymbol {M} \rangle),
\end{eqnarray}
where the $g_\mathrm{J}$ is the Land$\acute {e}$ $g$-factor(=4/5) and $\mu_\mathrm{B}$ is the Bohr magneton. $H_\mathrm{CEF}$ is the CEF Hamiltonian for the subspace of the $J=4$ multiplet in the $O_\mathrm{h}$ group \cite{CEF}, and can be written as 
\begin{equation}
H_\mathrm{CEF}=W[x\frac{O_4^0-5O_4^4}{60}+(1-|x|)\frac{O_6^0-21O_6^4}{1260}],
\end{equation}
where the parameter $W$ characterizes the overall strength of the CEF potential at the Pr site and $x$ parameterizes the relative strength of the sixth-order Stevens' operators. $n$ is a coefficient for the mean field  magnetic exchange interaction.

Bauer \textit{et al.} proposed two CEF energy level schemes to explain the observed peak in the magnetic susceptibility $\chi(T)$ at 3.5~K \cite{Bauer02}. Parameters of the CEF interaction for the two schemes (referred to as schemes A and B) are shown in Table I. In Table II the energy levels for these schemes are given. An important difference between these two level schemes is the ground state. The ground state in scheme A is a non-Kramers $\Gamma_3$ doublet, whereas that in scheme B a $\Gamma_1$ singlet.  The first excited state in both cases is the same magnetic $\Gamma_5$ triplet, located at 11~K and 6~K for schemes A and B, respectively.  Since the second and third excited levels in both schemes have much higher energies, the low-temperature and low-field properties are dominated by the low-lying two levels.

\begin{table}
\caption{Parameters of the CEF interaction and the magnetic molecular field coefficient for the two different CEF level schemes.}
\begin{center}
\begin{tabular}{cccc} \hline
&$W$ (K) & $x$ & $n$ (g/emu)\\ \hline
scheme A & -5.44 & -0.72 & -2300\\
scheme B & 1.85 & 0.5 & -6000\\ \hline
\end{tabular}
\end{center}
\end{table}
\begin{table}
\caption{Energy levels for the scheme A and B}
\begin{center}
\begin{tabular}{cc} \hline
scheme A & scheme B\\ \hline
$\Gamma_\mathrm{1}$(313~K)& $\Gamma_\mathrm{3}$(111~K) \\
$\Gamma_\mathrm{4}$(130~K)& $\Gamma_\mathrm{4}$(65~K) \\
$\Gamma_\mathrm{5}$(11~K)& $\Gamma_\mathrm{5}$(6~K) \\
$\Gamma_\mathrm{3}$(0~K)& $\Gamma_\mathrm{1}$(0~K) \\ \hline
\end{tabular}
\end{center}
\end{table}
Let us first discuss the present magnetic susceptibility data $\chi(T)$. In a low field limit, the $4f$ part of the susceptibility obeys 
\begin{equation}
\frac{1}{\chi_{4f}}=\frac{1}{\chi_\mathrm{CEF}}-n,
\end{equation}
where $\chi_\mathrm{CEF}$ denotes the single-ion magnetic susceptibility.
\begin{figure} 
\begin{center}
\includegraphics[scale=0.45]{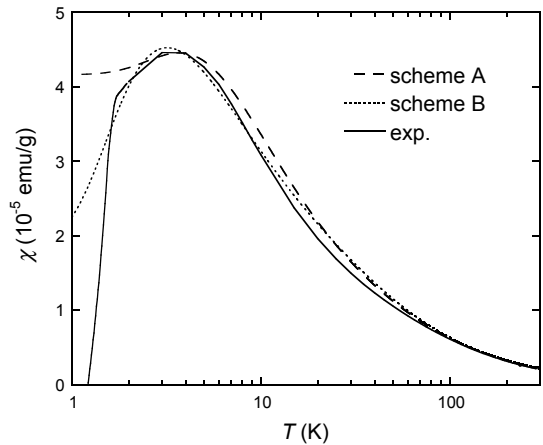}
\end{center}
\caption{Magnetic susceptibility of \pos\ for $H \parallel [100]$. The dashed and dotted lines are the calculated $\chi(T)$ for scheme A and B, respectively. Experimental data at $H=2$~kOe (solid line) are also shown for comparison.}
\end{figure}
The calculated $\chi_{4f}(T)$ for schemes A and B are shown in Fig.~8. The coefficient $n$ was determined so that $\chi_{4f}$ best reproduces the result at 2~kOe (solid line). The obtained values of $n$ for two schemes are shown in Table I. A main difference between the two calculated $\chi_{4f}$ is in the low-temperature behavior below 4~K. The $\chi_{4f}(T)$ curve for scheme A becomes nearly constant just after showing a weak peak at 4~K. On the other hand, $\chi_{4f}(T)$ for the scheme B strongly reduces upon cooling below 4~K, because of a difference in the Van Vleck contribution between the ground state and first excited state. Clearly, the $\chi_{4f}$ curve for scheme B is in better agreement with the experimental result, which continues decreasing down to $T_\mathrm{c}=1.5$~K. Of course, we cannot judge the ground state of
PrOs$_4$Sb$_{12}$ from this fact alone, since the hybridization effect cannot be neglected in this system and the simple localized electron model might not be applicable at such a low-temperature and low-field region.

\begin{figure}
\begin{center}
\includegraphics[scale=0.45]{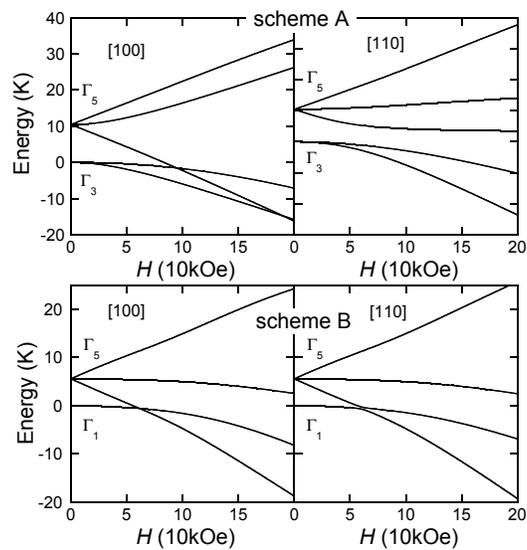}
\end{center}
\caption{Field dependence of the two lowest energy levels. Upper panels show the Zeeman effect for $\Gamma_3$ ground state model (scheme A), whereas the lower ones are those for the $\Gamma_1$ ground state model (scheme B).}
\end{figure}
\begin{figure} 
\begin{center}
\includegraphics[scale=0.45]{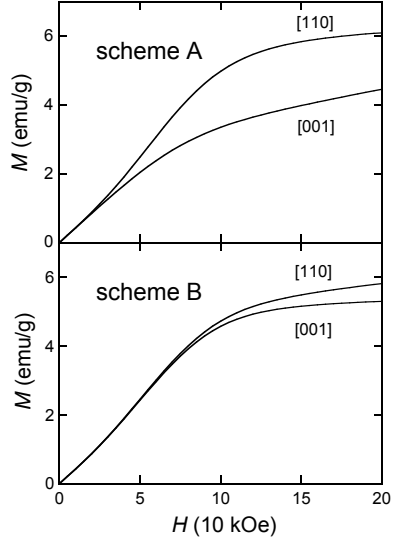}
\end{center}
\caption{Magnetization curves $M(H)$  at the paramagnetic state calculated at $T=2$~K for the $\Gamma_3$ (scheme A) and  the $\Gamma_1$ (scheme B) ground state models.}
\end{figure}
We next calculate the Zeeman splitting of the CEF levels for the two orientations [100] and [110] by using the parameters in Table I, and the results for the relevant low-lying levels are given in Fig.~9. In scheme A, the Zeeman effect is anisotropic. Accordingly, the calculated magnetization curves are strongly anisotropic, as shown in Fig.~10. It should be noticed that a level crossing in the ground state appears at 190~kOe only for $H \parallel [100]$, as shown in Fig.~9. 
On the contrary, the field dependence of the CEF levels in scheme B is almost isotropic, leading to a nearly isotropic magnetization curves as shown in Fig.~10.  It should be emphasized that a level crossing occurs at 60~kOe for both orientations, though the crossing for [110] is weakly of repulsive type.

\begin{figure} 
\begin{center}
\includegraphics[scale=0.45]{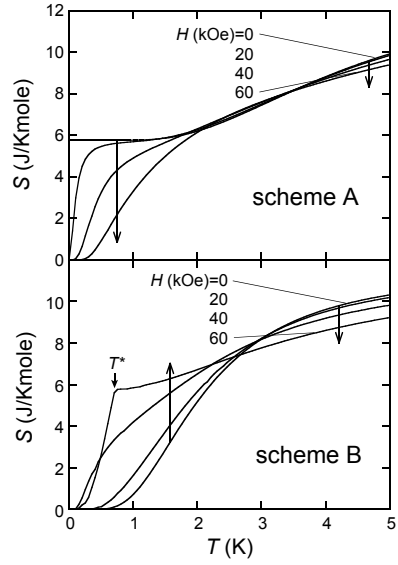}
\end{center}
\caption{Temperature dependence of the calculated entropy $S(T)$ 
at several fields along the [001] direction for the $\Gamma_3$ (scheme A) and  the $\Gamma_1$ (scheme B) ground state models.}
\end{figure}
Aoki \textit{et al.} measured the specific heat of \pos\ in magnetic fields and obtained the entropy $S$ as a function of $T$ for various fields \cite{Aoki02}. They revealed that the entropy increases with $H$ at low temperatures below $\sim 3$~K. The observation is rather unusual because the entropy is normally a decreasing function of magnetic field. They argued that the observed decrease in the entropy by the field is related to the CEF level crossing expected in scheme B. Here we extend the mean-field Hamiltonian (1) to a two-sublattice model to treat the field-induced ordering, and calculate the temperature dependence of the entropy $S(T)$ at various fields below 60~kOe for schemes A and B (Fig.~11). 
The calculated $S(T)$ for scheme A decreases in the whole temperature range as the field increases, whereas for scheme B the $S(T)$ value below  $\sim 2$~K increases with the field, which is in agreement with the experimental results by Aoki \textit{et al} \cite{Aoki02}. The kink at $T^*$=0.7~K in the $S(T)$ curve for 60~kOe is due to an AF magnetic transition caused around the level crossing. It can be shown that the field-induced AF ordering appears for all the field directions because of the nearly isotropic Zeeman effect. An AFQ transition is equally possible if we include an AFQ interaction in eq.~(1).   
On the other hand, it is obvious that no AF ordering can be expected
for the scheme A (upper panel of  Fig.~11) with the given weak AF
interaction, since the CEF ground state is non-magnetic and the level
crossing occurs only for $H \parallel [100]$ at a very high field ($\sim200$~kOe). Even if we
take a quadrupolar interaction into account, it is very hard to explain the
field-induced ordered phase (cf. Fig.~7) with the $\Gamma_3$ ground state model, so long
as we stay within the localized electron picture. In this situation, an AFQ
phase should appear from \textit{zero field} because the CEF ground state
itself possesses the quadrupolar degree of freedom. This AFQ phase may
exhibit a reentrant behavior like the case in PrPb$_3$ \cite{Tayama01}, but never exhibits a field-induced ordering like the one shown in Fig.~7.

\section{Discussion}
In the preceding section, we showed within the localized electron picture
that the magnetic properties of \pos\ are better explained by the
$\Gamma_1$ ground state model. In particular, there is a serious difficulty
in the $\Gamma_3$ ground state model on explaining the field-induced
ordering. Of course, we cannot go too far with the localized electron
picture because the $4f$ electrons in this compound are considered to be
delocalized at low temperatures due to hybridization with conduction electrons. If the
delocalization effect is strong enough to destroy the quadrupolar ordering
at low fields, then a completely different situation might appear in the
$\Gamma_3$ ground state model: a scenario that a quadrupolar ordering is restored in strong field by suppression of the delocalization effect.
A magnetic analogue of this scenario might be a field-induced antiferromagnetic phase transition in a system where antiferromagnetic interactions among localized magnetic moments are competing with the Kondo effect. The fact that such a transition has not been observed yet in any Ce-based hf compounds suggests that the aforementioned scenario for $\Gamma_3$ is unlikely.

A remaining problem with respect to the $\Gamma_1$ ground state model is how the heavy-fermion state is formed at low temperatures. Concerning this, there are several theoretical studies, which attempt to explain the Fermi liquid state with heavy quasiparticles in the $f^2$ configuration with a singlet CEF ground state by taking the CEF excited states into consideration \cite{Ikeda97,Watanabe97}. Actually, many U-based-hf compounds (e.g. UPt$_3$, and URu$_2$Si$_2$),  which are likely to be in the $f^2$ configuration, is believed to possess a singlet CEF ground state. Accordingly, formation of the hf state of \pos\ might be understood by the $\Gamma_1$ ground state model.

As for the double sc transition, at present there is no clear-cut evidence that this phenomenon is intrinsic. Since the phase boundaries $T_\mathrm{c1}(H)$ and $T_\mathrm{c2}(H)$ are likely to be scaled to each other (Fig.~7),  there remains a possibility of a phase separation in spite of the high quality of the sample: a small fraction of the volume first undergoes a sc transition at $T_\mathrm{c1}$, and at slightly lower temperature $T_\mathrm{c2}$ the whole volume of the sample changes into the sc state. To eliminate this scenario, further experimental efforts will be needed.

\section{Summary}
We measured the dc magnetization of \pos\ in low temperatures down to 60~mK. The presence of the double sc transition, which was first observed in the specific heat measurements at zero field, is confirmed in fields up to 14~kOe. The field dependence of $T_\mathrm{c2}$ is qualitatively the same as that of $T_\mathrm{c1}$. We also observed phase A not only for [100] direction but also [110] and [111] directions. The anisotropy of the A-phase transition temperature is $T_\mathrm{A}^{[100]} > T_\mathrm{A}^{[111]} >T_\mathrm{A}^{[110]}$. For $H \parallel [110]$ two additional transitions are found inside phase A. Comparison between the experimental results and the calculations on the basis of the CEF model suggests that the CEF ground state of Pr$^{3+}$ is a $\Gamma_1$ singlet, in contrast to the previous report \cite{Bauer02,Maple02}. 

\section*{Acknowledgments}
We are grateful to Y. Matsuda and K. Machida for valuable discussions. The present work was supported by Grant-in-Aid for Scientific Research.


\begin{thebibliography}{29}
\bibitem{CEF} The correct point group symmetry for \pos\ is $T_\mathrm{h}$ as reported in ref.~\cite{Takegahara01}. In the $T_\mathrm{h}$ group an additional sixth order term appears in the CEF Hamiltonian (3). This term mixes the wave function of the $\Gamma_4$ and $\Gamma_5$ triplets with each other. If this term is predominant, the CEF ground state becomes magnetic. This is inconsistent with the experimental results. Thus, the additional term would be small in this case.  Here we use the CEF Hamiltonian for the $O_\mathrm{h}$ group for simplicity.
\end{thebibliography}
\end{document}